\documentclass[aps,prd,preprint,superscriptaddress]{revtex4-1}

\usepackage{graphicx}
\usepackage{booktabs}
\usepackage{hyperref}
\usepackage{amssymb,amsmath,amsfonts}
\usepackage{color}
\usepackage[utf8]{inputenc}

\begin{document}


\title{Volume dependence of baryon number cumulants and their ratios}

\author{G\'abor A. Alm\'asi}
\affiliation{Gesellschaft f\"{u}r Schwerionenforschung, GSI, D-64291 Darmstadt, Germany}
\affiliation{Technische Universit\"{a}t Darmstadt, D-64289 Darmstadt, Germany}

\author{Robert D. Pisarski}
\affiliation{Department of Physics, Brookhaven National Laboratory, Upton, NY 11973}
\affiliation{RIKEN/BNL Research Center, Brookhaven National Laboratory, Upton, NY 11973}

\author{Vladimir V. Skokov}
\affiliation{RIKEN/BNL Research Center, Brookhaven National Laboratory, Upton, NY 11973}

\begin{abstract}
We explore the influence of finite volume effects on baryon number fluctuations in a
non-perturbative chiral model. In order to account for soft modes, 
we use the functional renormalization group in a finite volume, using
a smooth regulator function in momentum space.
We compare the results for a smooth regulator with those for a sharp (or Litim) regulator,
and show that in a finite volume, the latter produces spurious artifacts.
In a finite volume there are only apparent critical points, about which we
compute the ratio of the  fourth to the second 
order cumulant of quark number fluctuations.
When the volume is sufficiently small the system has two apparent critical points;
as the system size decreases, the location of the apparent critical point
can move to higher
temperature and lower chemical potential.
\end{abstract}

\maketitle

\section{Introduction}
Experiments with ultrarelativistic
heavy-ion collisions at RHIC and LHC explore the phase structure of 
Quantum ChromoDynamics (QCD) at nonzero temperature and density, and
so probe the phase transitions associated with
deconfinement and the restoration of chiral symmetry.
Two of the most promising observables are the
fluctuations of the net baryon number and electric
charge. The cumulants and related quantities  (see e.g. Ref.~\cite{Bzdak:2016sxg})   
of these fluctuations may provide experimental evidence for a
chiral critical endpoint~\cite{Stephanov:1998dy,Stephanov:1999zu,Stephanov:2011pb} 
or chirally inhomogenous phases.

The interest in the analysis of cumulants is not restricted 
only to high baryon densities.  As was pointed out 
in Ref.~\cite{Friman:2011pf,Skokov:2011yb}, 
higher order cumulants  reflect the underlying $O(4)$
critical dynamics, as cumulants of higher order
are driven to negative values at temperatures close to 
that for a phase transition. 
This may lead to  a strong suppression of
the higher order cumulants, and help to identify
the chiral crossover experimentally.

The STAR collaboration has measured fluctuations in the 
net proton number, as a proxy for the net baryon number,
and demonstrated that the
kurtosis depends non-monotonically on the collision energy~\cite{Adamczyk:2013dal}. 
This may serve as a strong indication of the chiral critical endpoint.  

However, there are many other effects besides the critical dynamics
which might be important in the interpretation of the data. 
Those include the conservation of baryon number~\cite{Bzdak:2012an}, 
corrections for efficiency in the detectors~\cite{Bzdak:2013pha}, 
hadronic  rescattering~\cite{Kitazawa:2011wh},  
non-equilibrium effects~\cite{Mukherjee:2015swa,Mukherjee:2016kyu},  and 
finally volume fluctuations~\cite{Skokov:2012ds,Braun-Munzinger:2016yjz, Palhares:2009tf, *Fraga:2011hi,*Palhares:2012zz, *Hippert:2015rwa}.
The latter are important due to a finite size of a domain
passing through the critical region during the evolution of the fireball.
Usually one tries to minimize the effects of fluctuations in the volume by considering the
ratios of cumulants. As we describe in the main text, 
in such ratios the explicit dependence on the volume cancels 
out, making the analysis of volume fluctuations trivial.
However, we show that the implicit dependence on the 
volume might be very strong if the characteristic system size is below 
$\sim 5$ fm. 

In this paper we compute using
the functional renormalization group (FRG) in a Quark-Meson (QM) model.
In the next section we formulate the 
FRG approach to the QM model in a finite volume and stress 
the importance of using a smooth cut-off function in momentum space for the FRG. In Sec.~\ref{Sec:T=0}, 
we show how the chiral order parameter depends on the size of the system.
We perform the calculation in a box with both isotropic and anisotropic 
dimensions. In Sec.~\ref{Sec:ACEP} we find the location of apparent 
critical points and trace their dependence on the system size.
Finally,  in Sec.~\ref{Sec:Cumulants} we compute the cumulants of 
quark number fluctuations for different system sizes and anisotropies.


\section{Functional renormalization group for chiral models in finite volume}

In this paper we use the quark-meson model as a 
realization of the chiral symmetry in QCD at low energies.
The quark-meson model consists of a 
$O(4)$ multiplet of mesons,
$\phi=(\sigma,\vec{\pi})$, 
coupled to quark fields $q$ through a Yukawa-type coupling, $y$. The
Lagrangian density is given by
\begin{gather}
 \mathcal{L}=\bar{q}[i\gamma_\mu \partial^\mu - y(\sigma + i\gamma_5
  \vec{\tau}\cdot\vec{\pi}) ]q +\frac{1}{2}(\partial_\mu
  \sigma)^2  +\frac{1}{2}(\partial_\mu \vec{\pi})^2 \nonumber -U(\sigma,\vec{\pi}),
\end{gather}
where $U(\sigma,\vec{\pi})$ denotes the mesonic potential,
\begin{equation}
U(\sigma,\vec{\pi}) = \frac{1}{2}m^2 \phi^2 + \frac{\lambda}{4}\left(\phi^2\right)^2 - h\sigma \; ,
\end{equation}
$\phi^2 = \sigma^2 + \vec{\pi}^2$.
For $m^2<0$ and $h=0$,  the $O(4)$ symmetry of the potential  is
spontaneously broken to $O(3)$, resulting in a  non-vanishing value of
the vacuum scalar condensate $\langle \sigma \rangle$ and a non-zero quark mass.
The last term,  $h=f_\pi m_\pi^2$,   breaks the chiral
symmetry explicitly and yields a nonzero pion mass.

In order to formulate a non-perturbative thermodynamics in the QM model 
we adopt a method based on the functional renormalization group (FRG). 
The FRG is based on an infrared regularization with the momentum scale parameter, $k$,  
where the full propagator is derived from a 
corresponding effective action, $\Gamma_k$.

For an infinite volume, in the Local Potential Approximation \cite{Berges:2000ew}
the FRG equation for the quark-meson model is
\begin{equation}
\label{eq:gen_floweq_fin}
\partial_k \Omega = \frac{1}{4} \int \frac{d^3q}{(2\pi)^3}
	\left( \frac{1+2n_B(E_{\sigma})}{E_{\sigma}} + 3\frac{1+2n_B(E_{\pi})}{E_{\pi}} -2\nu_q\frac{1-n_F(E_q)-\bar{n}_F(E_q)}{E_q}\right) \partial_k R_k(q)\, ,
\end{equation}
where $\nu_q = 2 N_c N_f$ is the fermion degeneracy factor, $R_k(q)$ is the regulator function,
\begin{equation}
	n_B(E)=\frac{1}{e^{E\beta}-1},\quad n_F(E)=\frac{1}{e^{(E-\mu)\beta}+1},
	\quad \bar{n}_F(E)=\frac{1}{e^{(E+\mu)\beta}+1}	
\end{equation}
are the Bose-Einstein and Fermi-Dirac distribution functions with the quasi-particle energies defined as
\begin{equation}
	E_{x} = \sqrt{m_x^2 + q^2 + R_k(q)}
\end{equation}
and $\beta=1/T$ is the inverse temperature. The masses for the quasi-particles are
\begin{align}
	m_\sigma^2 &= \frac{\partial^2 \Omega}{\partial \sigma^2}\,, \\  
	m_\pi^2 &= \frac{1}{\sigma} \frac{\partial \Omega}{\partial \sigma}\,, \\  
	m_q &= y \sigma .  
	\label{Eq:masses}
\end{align}
Here we assume the symmetry is broken in the $\sigma$ direction.

The solutions of the FRG flow equation determines $\Omega({k \to 0}, \sigma)$ at any  
possible value of $\sigma$. We are interested in the equilibrium value, 
which can be found by locating the minimum of the thermodynamic potential,  $\Omega({k \to 0}, \sigma)$. 
In the presence of the explicit symmetry breaking term, instead
of minimizing $\Omega({k \to 0}, \sigma)$, one minimizes $\Omega({k \to 0}, \sigma) - h \sigma$. 
Note that the symmetry breaking parameter $h$ does not enter the FRG evolution equation, which 
is solved  for any value of $\sigma$.

In a finite volume we consider periodic boundary conditions, and that means 
momentum integrals are replaced by summations. The general flow equation becomes
\begin{equation}
\label{eq:gen_floweq_fin_1}
	\partial_k \Omega = \frac{1}{4L^3} \sum_{n_x,n_y,n_z}
	\left( \frac{1+2n_B(E_{\sigma})}{E_{\sigma}} + 3\frac{1+2n_B(E_{\pi})}{E_{\pi}} -2\nu_q\frac{1-n_F(E_q)-\bar{n}_F(E_q)}{E_q}\right) \partial_k R_k(q).
\end{equation}
The external momentum $q$ is an appropriate function of the modes $q=q(\vec{n})$ 
defined by the boundary conditions. 
For the box of dimensions $L_x$, $L_y$, and $L_z$, with periodic boundary conditions we have 
\begin{equation}
	q^2 = \sum_{i=x,y,z} \left( \frac{2\pi n_i}{L_i} \right)^2\, .
\end{equation}
In this work, we do not restrict ourselves a symmetric box where $L_x=L_y=L_z=L$, but consider 
as well geometries with equal transversal extents, $L_x=L_y=L$, and
longer in $L_z = A L$. The mode summation in Eq.~\eqref{eq:gen_floweq_fin} 
is performed numerically by introducing a multiplicity of state 
function, as described in Appendix \ref{Sec:ModeSum}.
The numerical algorithms and the input parameters are detailed in Appendix \ref{Sec:Num}.

So far we have not defined the regulator function $R_k(q)$ that we use. 
Previously, regulator functions which are sharp in momentum space have been
used (see, {\it e.g.}, Refs.~\cite{Braun:2011iz,Tripolt:2013zfa}). The most popular one  is the  Litim regulator: 
\begin{equation}
	R_k^{\rm sharp}(q) = (k^2-q^2)\theta(k^2-q^2),
	\label{Eq:Sharp}
\end{equation}
where $\theta$ denotes the Heaviside step function, $\theta(x) = 0$ for $x < 0$ and $=1$ for $x >0$.

However, sharp cutoffs results in numerical artifacts.  As we show in the next section,
these include oscillations in the  order parameters and meson masses.
Mathematically, a sharp cutoff is awkward for the fact that the spatial
momenta are quantized in a finite volume, as discussed by Fister and Pawlowski~\cite{Fister:2015eca}.  
In this paper, we adopt an exponential cutoff 
\begin{equation}
	R_k(q) = \frac{q^2}{e^{q^2/k^2}-1}.
\end{equation}
Such an exponential clearly cuts off fluctuations above the momentum
scale $k$, but does so smoothly.

\section{Zero temperature and chemical potential}
\label{Sec:T=0}
In this section we compute
expectation value of the order parameter and its flow as a function of the cut
off momentum $k$. We compare our results, with a smooth, exponential regulator,
to those with a sharp cutoff, Eq.~\eqref{Eq:Sharp}.

\begin{figure}
\includegraphics[width=0.7\linewidth]{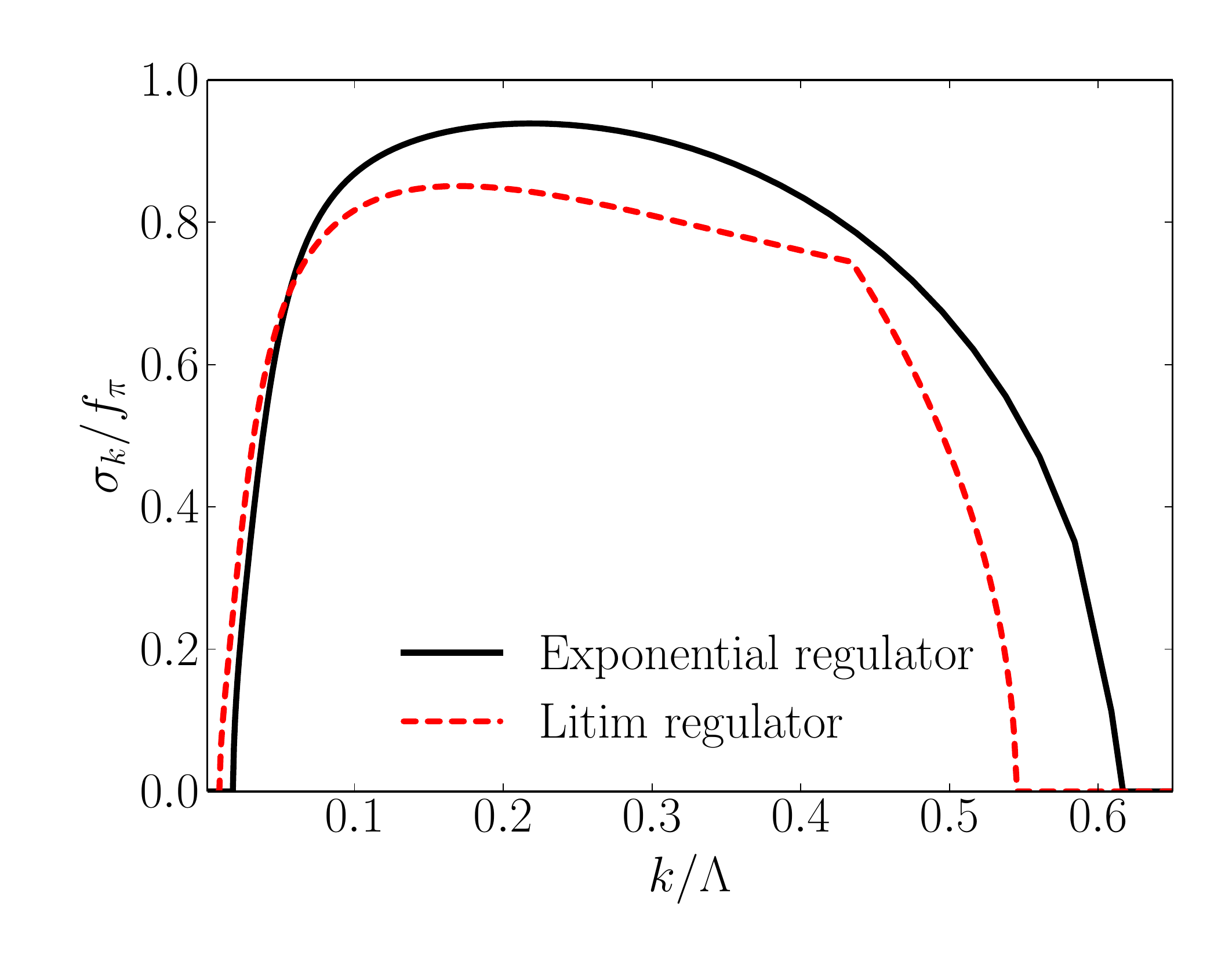}
\caption{The flow of the chiral condensate, $\sigma$, in the chiral limit as a function of the scale $k$ 
for the exponential and Litim regulators. The scale dependent condensate is obtained by 
minimizing the potential at each RG scale. The system size was chosen to be $L=3$ fm, at zero
temperature and chemical potential.}
\label{Fig:Flow}
\end{figure}

In Fig.~\ref{Fig:Flow} we show the flow of the expectation value 
of the order parameter, which is the location of the minimum of the potential, in the chiral limit. 
As expected in the IR limit, the expectation of the order parameter is zero,
demonstrating the absence of spontaneous symmetry breaking 
in a finite volume.  We discuss this point further in Appendix~\ref{Sec:ASB}.
We also show this figure to demonstrate that a sharp regulator in momentum
space produces non-analytic flow in the Functional Renormalization Group.

\begin{figure}
\includegraphics[width=0.7\linewidth]{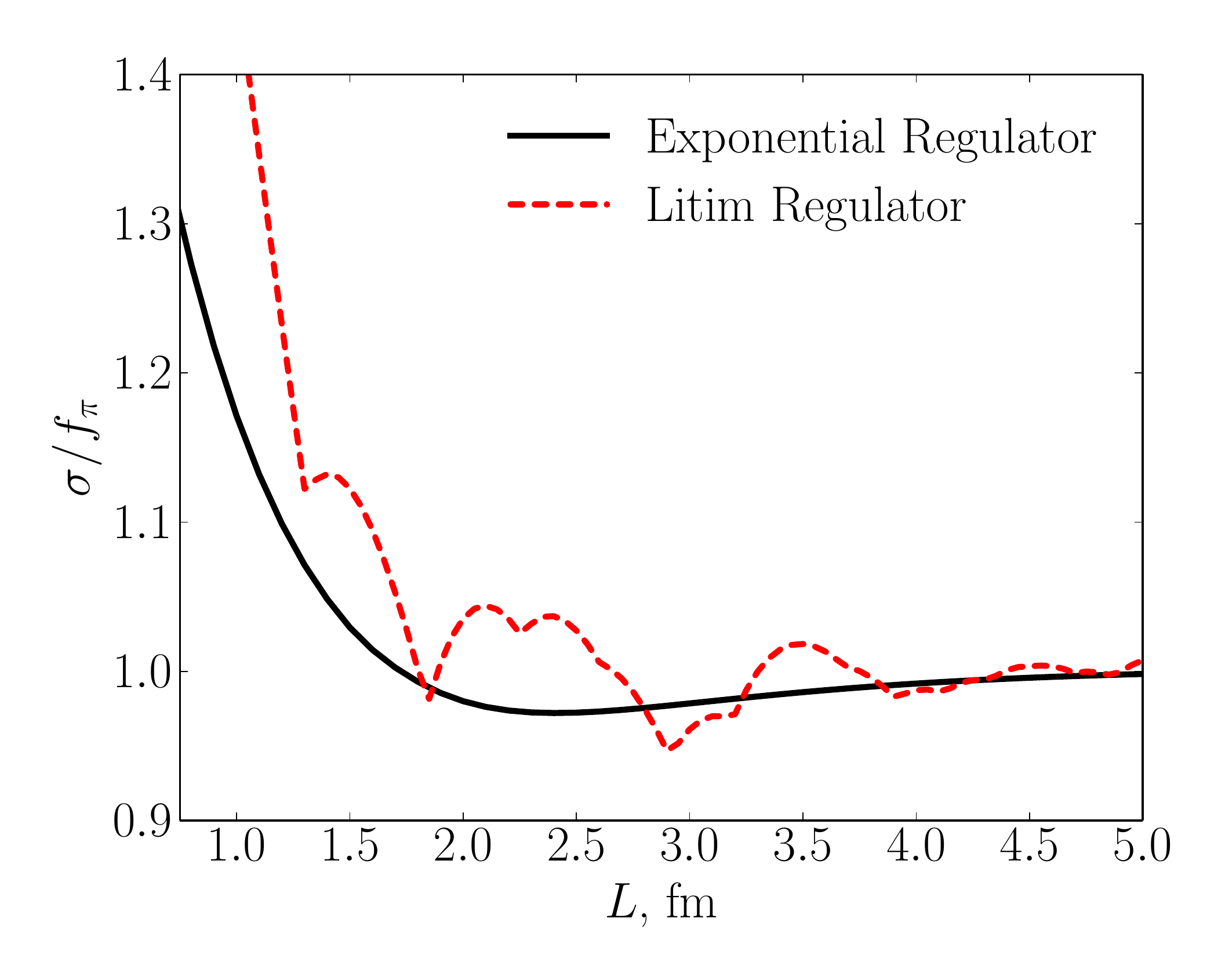}
\caption{The chiral condensate, $\sigma$, as a function of the system size, $L$,  
at zero temperature and chemical potential for the exponential and the Litim regulators.}
\label{Fig:SigmaBC}
\end{figure}

These artifacts become more prominent when we plot the dependence of the 
order parameter on the size of the system. In Fig.~\ref{Fig:SigmaBC} we perform the calculations at a
physical pion mass. We have checked our computations analytically
in the limits of small and large volumes, $L\to0$ and $L\to\infty$, in Appendix~\ref{Sec:Limits}.  

These artifacts are elementary to understand.  In a finite volume we uniformly
take periodic boundary conditions, so that each momentum is a multiple  of $2 \pi/L$.
With a sharp cutoff in momentum space, then, the momenta included
by the Functional Renormalization Group jumps whenever $2 \pi/L$ crosses that cutoff.
With a smooth cutoff, the effects of high momenta are automatically included, but vanish
smoothly, and so do not produce any artificial discontinuities.

\begin{figure}
\includegraphics[width=0.7\linewidth]{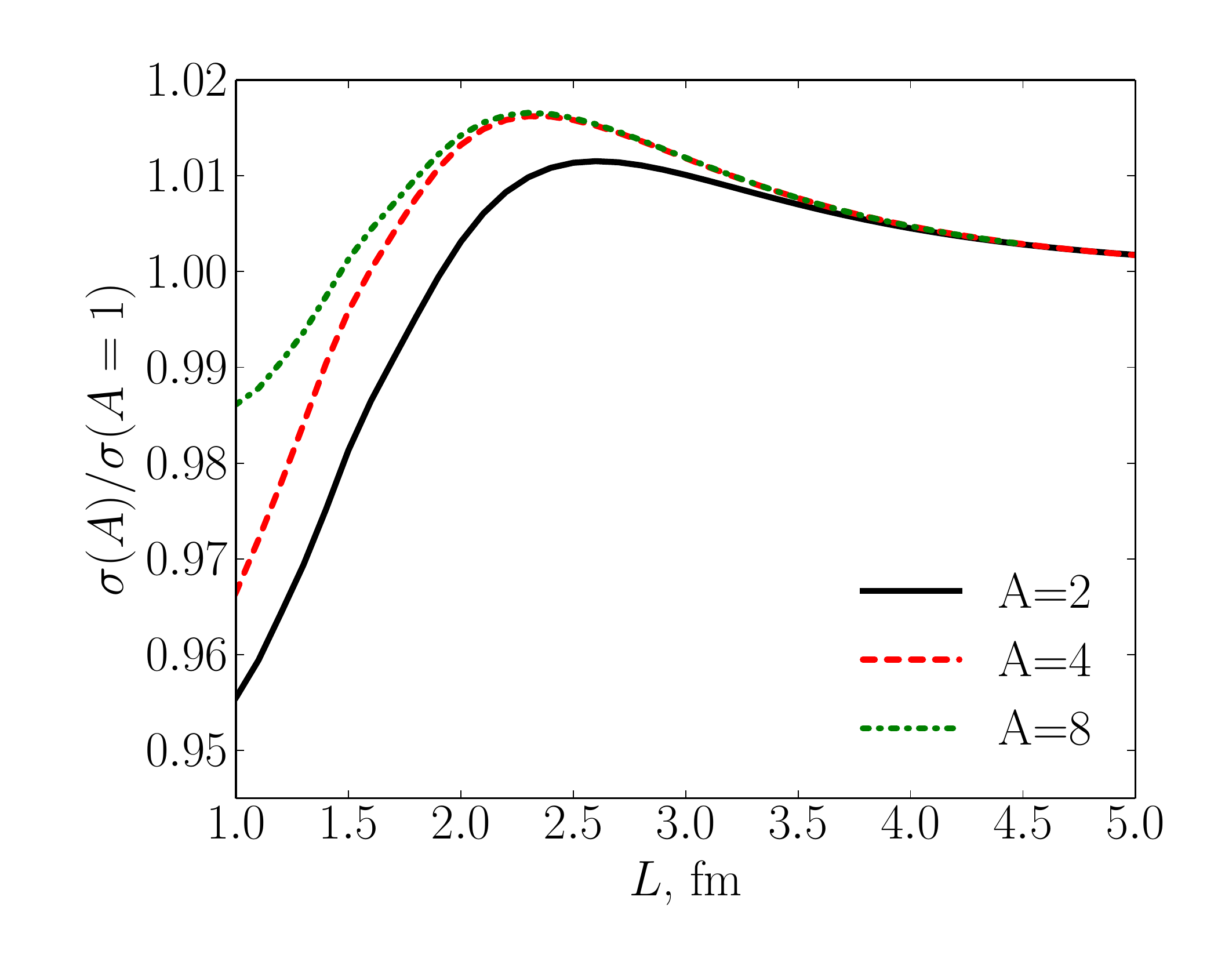}
\caption{
The  chiral condensate as a function of the system size, $L=L_x=L_y$ for the 
different  anisotropy parameter  $A \equiv L_{z}  / L_{x,y}$. The results are 
normalized by the corresponding values of the chiral condensate for the 
isotropic volume, $L_x=L_y=L_z=L$. Periodic boundary conditions are used.  
}
\label{Fig:SigmaA}
\end{figure}

In Fig.~\ref{Fig:SigmaA}, we show how the order parameter
depends upon the anisotropy parameter $A$.  The results differ for
small volume; as the volume increases, the curves approach that for
infinite volume, regardless of the value of the anisotropy. 
For any two given values of the anisotropy parameter $A_1$ and $A_2$,
such that $A_2>A_1$, the corresponding chiral condensates  $\sigma(A_1, L)$ and 
$\sigma(A_1, L)$ start to differ if the system size is $L A_1 < $8 fm.

\section{Location of apparent critical end point}
\label{Sec:ACEP}
We consider systems in which there is a true critical point in infinite
volume.  In finite volume, instead there is an 
{\it apparent} critical point (ACP).   There is some degree of 
arbitrariness in how one defines an apparent critical point.
We define the position of the apparent critical point 
from the maximum in the corresponding chiral susceptibility, 
which is equivalent to the minimum in the sigma mass, $m_\sigma$. 
We stress, however, that unlike the case of infinite volume, that in finite
volume other definitions will give different positions for the apparent
critical point.

\begin{figure}
\includegraphics[width=0.7\linewidth]{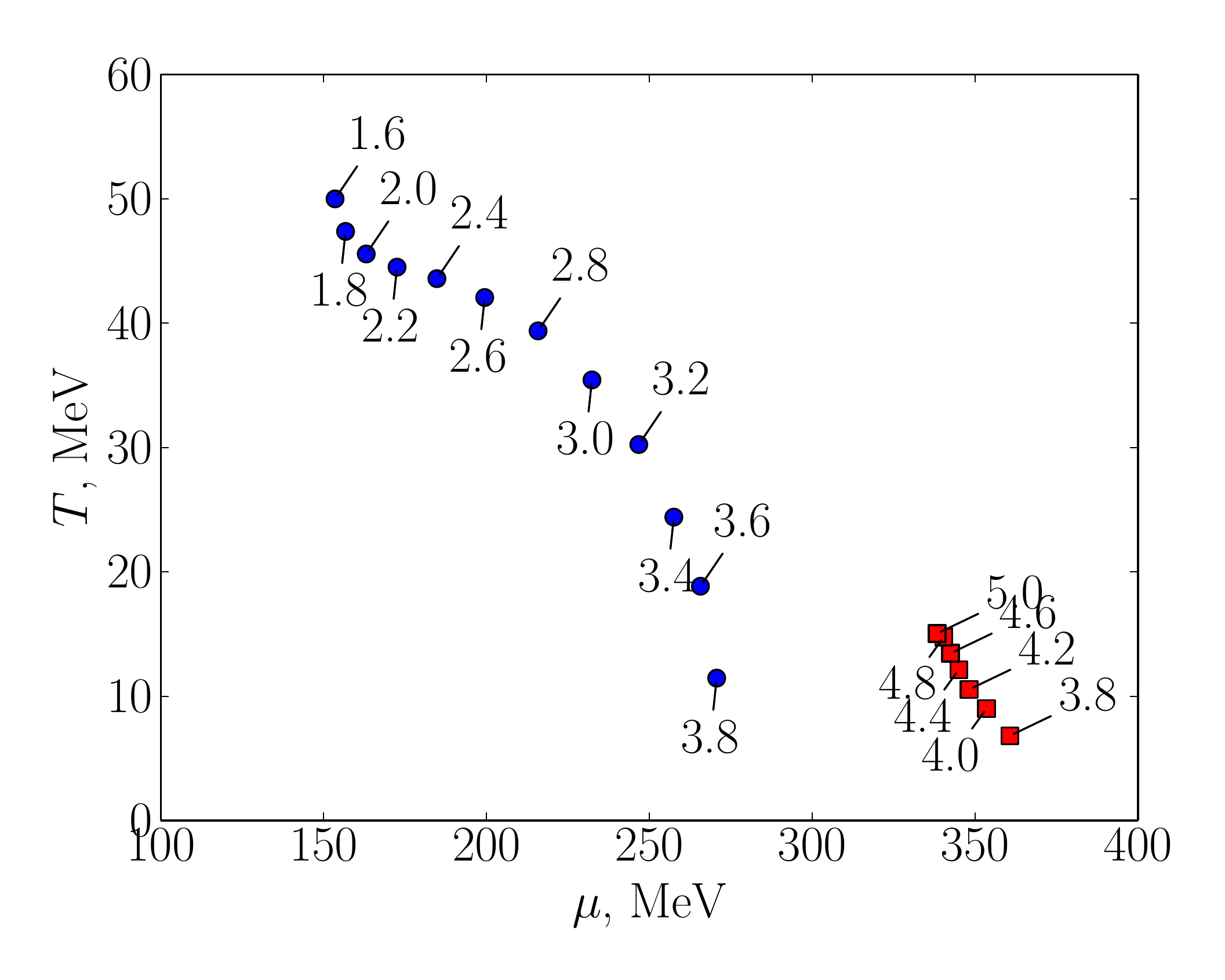}
\caption{
	The location of the apparent critical points (ACP) as a function of 
        the system size, $L$.
		Due to 
	the numerical difficulties we were not able to resolve ACPs
	at temperatures below 5 MeV. The red points continuously 
	approach the true critical point in the limit of infinite volume,
        which is already well approximated by $L=5$ fm.  
}
\label{Fig:ACEP}
\end{figure}

With our definition, we show that at some
intermediate system size, the system has {\it two} apparent
critical points, located at different
values of $T$ and $\mu$.   One of the apparent
critical points, which we call ACP I, approaches
the true critical point in the limit of infinite volume; we show that for
the ACP I, it
approaches the zero temperature axis as the volume decreases.  The second
apparent critical point, which we call ACP II, appears near the zero temperature
axis, and evolves to {\it higher} temperature as the volume decreases.
The location of the two apparent critical points is depicted in Fig.~\ref{Fig:ACEP}.
The emergence of a second apparent critical point influences the cumulants of 
baryon number, and is studied in the next section. 

To grasp the essence of the behavior of the critical point, we carried out
a mean-field calculation by omitting the bosonic contribution. A recent study suggests that the apparent critical point within this approximation shows a qualitatively similar behavior \cite{Juricic:2016tpt}. We refitted the parameters to reproduce $m_q = 335\; \textrm{MeV}$ and $m_{\sigma}=500\; \textrm{MeV}$.
For transparency, we chose a slightly smaller sigma mass compared to our 
calculations. In this case the first-order phase 
transition occurs at slightly smaller chemical potential, and so the
minimum of the potential at $\sigma=f_{\pi}$ is not influenced by finite 
density effects at the relevant chemical potentials.

In mean-field calculations one drops bosonic fluctuations, and 
there appears to be a second order chiral phase 
transition even in finite volume. For simplicity, we consider the 
the chiral limit at zero temperature. For each volume, at some 
intermediate chemical potential the system goes from the ground
state at $\sigma=f_{\pi}$ to a chirally restored phase at $\sigma=0$ through
a first order transition.  We assume that where this transition happens
on the $T=0$ axis is related to the location of the critical end-point
in the plane of temperature and chemical potential.

\begin{figure}
\includegraphics[width=0.7\linewidth]{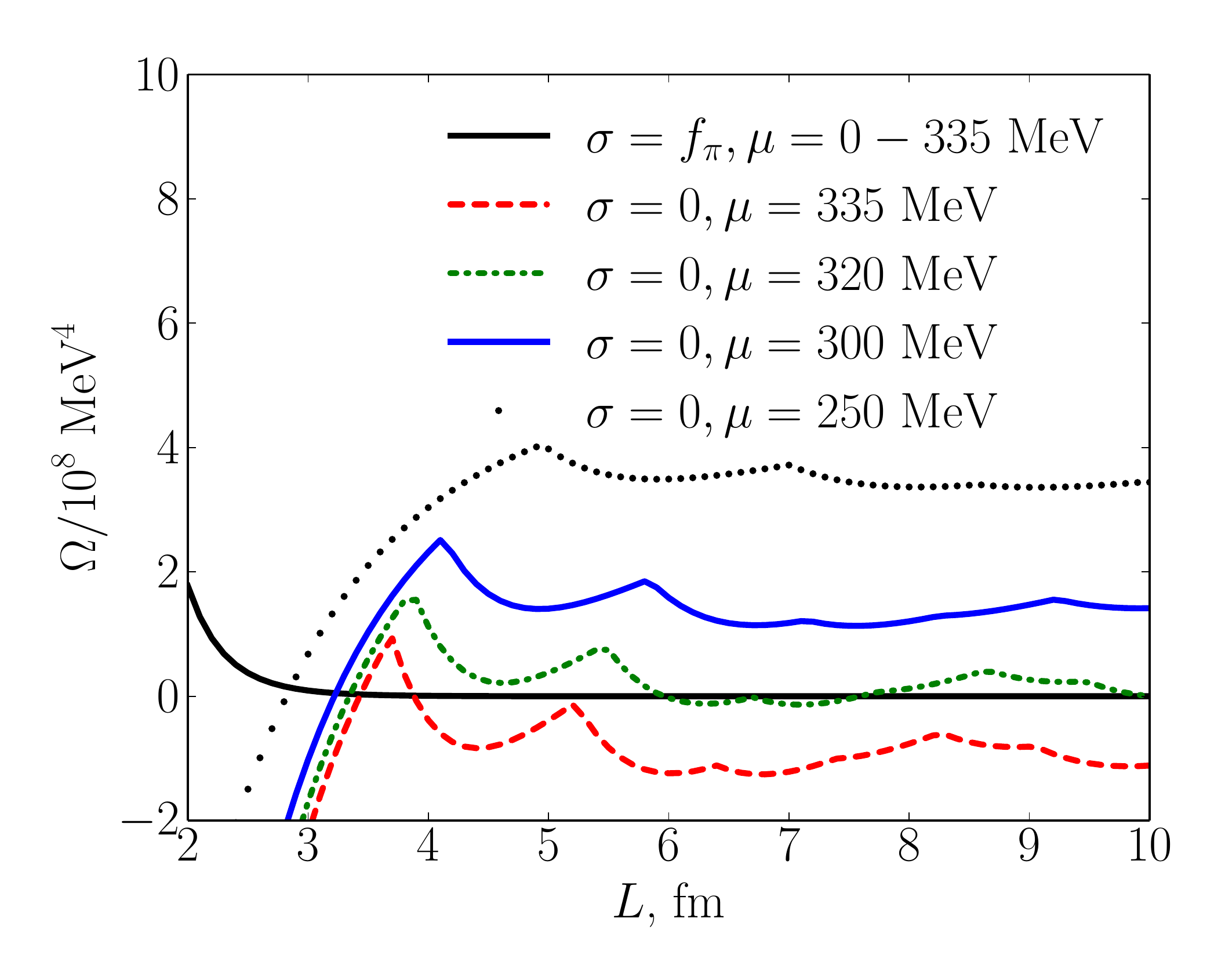}
\caption{
The potential at $\sigma=f_{\pi}$ and  $\sigma=0$
as a function of the system size for different chemical potentials. 
}
\label{Fig:ACEPT0}
\end{figure}

On the axis where $T=0$, the phase transition occurs when the condition
\begin{equation}
	\label{eq:transition}
	\Omega(\mu,L,\sigma=f_{\pi}) = \Omega(\mu,L,\sigma=0)
\end{equation}
is fulfilled. For chemical potentials $\mu < g f_{\pi}$, the left-hand side is 
independent of $\mu$.  After subtracting the value in infinite volume, 
\begin{equation}
	\Omega(\mu<g f_{\pi},L,\sigma=f_{\pi})-\Omega(0,\infty,f_{\pi}) = \nu_q \left(\frac{1}{L^3}\sum_{n_x,n_y,n_z} (E_1-E_2) - \frac{1}{(2\pi)^3}\int d^3p\; (E_1-E_2) \right),
\end{equation}
with
\begin{equation}
	E_1= \sqrt{g^2f_{\pi}^2 + q^2+R_{\Lambda}(q)},\quad E_2= \sqrt{g^2f_{\pi}^2 + q^2},\quad q =\frac{2\pi}{L} \sqrt{n_x^2+n_y^2+n_z^2}.
\end{equation}
This is depicted by the black, solid line in Fig. \ref{Fig:ACEPT0}. 
The right-hand side of Eq.~\eqref{eq:transition} depends on the chemical potential and is given by
\begin{equation}
	\label{Eq:OmegaMFT0}
	\Omega(\mu,L,0) = \frac{\nu_q}{L^3} \sum_{n_x,n_y,n_z} (\sqrt{q^2+R_{\Lambda}(q)}-q-(\mu-q)\theta(\mu-q)),
\end{equation}
where $\theta$ denotes again the Heaviside step function.
Let us consider the finite 
density part of this function at constant chemical potential as the size of
the system changes.
It is expected that the finite density part ($\mu$-dependent term in Eq.~\eqref{Eq:OmegaMFT0}) will 
be affected by  
finite-volume effects stronger than the vacuum part, since we only probe modes 
up to the Fermi surface, and in small volumes they are few in number.
The finite density part contributes  with a negative sign, so as it gets larger, 
the value of the potential decreases driving a phase transition.

As the volume decreases, the contribution of each mode is $\sim 1/L^3$, and so 
the total increases.  On the other hand, as $L$ decreases the momentum of each
mode goes up, $\sim 2 \pi/L$, so in all fewer modes fall 
below the Fermi momentum $q_f=\mu$. In total, there is a balance between these two effects,
so that at large $L$ there is an oscillatory behavior, as shown in Fig. \ref{Fig:ACEPT0}.
If the volume is very small, say below $L=3\;\mathrm{fm}$, only the
zero mode contributes. 
 This enhances the quark contribution to the potential at nonzero density
at small volume, and triggers a first order phase 
transition at lower values of the chemical potential.  
This is illustrated in Fig. \ref{Fig:ACEPT0},
where the potential of the $\sigma=f_{\pi}$ solution is compared to the $\sigma=0$ 
solution at different chemical potentials. When the two curves cross, there is a first 
order phase transition in the given volume at the corresponding chemical potential. 
In relatively high volumes, i.e. above $L=4\; \mathrm{fm}$, the phase transition 
occurs in the range $\mu=320-335 \; \mathrm{MeV}$. Its location as the function of system 
size is not monotonous due to oscillations. At low volumes, i.e.
$L <3.5\; \mathrm{fm}$, only the zero mode contributes to the finite density part, 
and the phase transition moves down to very low chemical potential:
at $L=3\; \mathrm{fm}$ the chemical potential is $\mu=250\; \textrm{MeV}$. 
As the temperature is turned on,
the transition line is expected to reach higher temperatures, since the phase 
transition at $T=0$ is strengthened by the sudden change in the zero mode contribution to
$\Omega(\sigma)$ at $\mu=g\sigma$.

This brief analysis suggests that the behavior of the ACP I is difficult to understand,
as its location may show oscillatory behavior. At small volumes,
the position of the apparent critical point is expected to move to very low 
chemical potentials, and its temperature is also expected to increase. This is in 
agreement with our findings about ACP II.

\section{Cumulants} 
\label{Sec:Cumulants}
In this section we discuss the dependence of the cumulants
of baryon number fluctuations on the size of the system.
In particular, we consider the ratio of 
the fourth to the second order cumulant of quark number fluctuations.
Up to an overall factor of $1/9$, this corresponds to the same ratio for
baryon number.
The second and fourth order cumulants for quark number are 
\begin{align}
	c_2 &= \langle (\delta N_q)^2 \rangle, \\ 
	c_4 &= \langle (\delta N_q)^4 \rangle - 3 \langle (\delta N_q)^2 \rangle^2  
	\label{Eq:c2andc4}
\end{align}
respectively, where $\delta N_q = N_q - \langle N_q \rangle$. 

In the limit of infinite volume a cumulant $c_n$ is proportional to the volume
times the susceptibility
\begin{equation}
	c_n = V T^3 \; \chi_n  \; ,
	\label{Eq:chi_c_relation}
\end{equation}
where 
\begin{equation}
	\chi_n = \frac{\partial^n  }{\partial (\mu/T)^n} \left( \frac{p}{T^4} \right) \; .
	\label{Eq:chi_n}
\end{equation}

Thus in infinite volume, it is natural to go from the experimentally observable cumulants
to the susceptibilities by taking their ratio,
\begin{equation}
	c_4/c_2 = \chi_4/\chi_2 . 
\end{equation}
In a finite volume, however, the factors of volume do {\it not} cancel.  As we demonstrated in the previous
section, the value of the chiral condensate depends upon the volume, and this
influences the position of any apparent critical point.

In our model, we derived flow equations for the density, which is closely related to $\chi_1$. Using numerical derivatives with respect to the chemical potential, we were able to extract $\chi_2$ and $\chi_4$. In order to get the correct high temperature behavior, we took into account quark contributions above the UV cutoff perturbatively. This is discussed in Appendix \ref{Sec:Pert}.

\begin{figure}
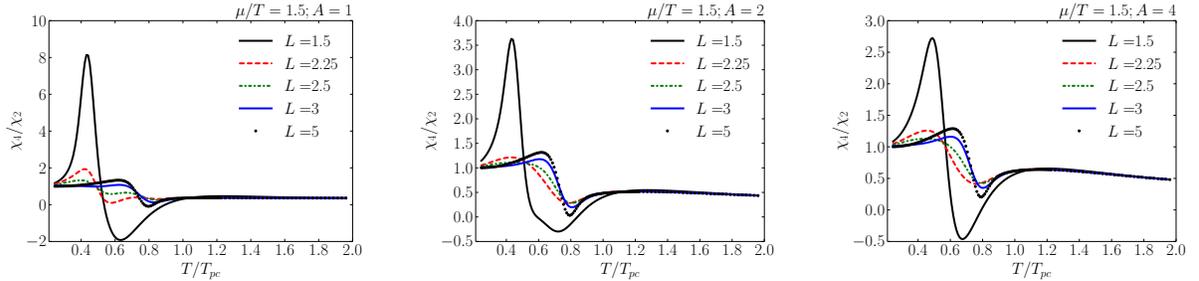

	\includegraphics[width=0.325\linewidth]{{{test_1.5_L}}}
	\includegraphics[width=0.325\linewidth]{{{test_1.5_as2_L}}}
	\includegraphics[width=0.325\linewidth]{{{test_1.5_as4_L}}}
\caption{
	The ratio of the fourth to the second order susceptibilities 
	as a function of temperature for different systems sizes 
	and the anisotropy parameter $A$; the results are computed at 
	$\mu/T=1.5$. 
}
\label{Fig:X1.5}
\end{figure}

In Figs.~\ref{Fig:X0},~\ref{Fig:X0.5},~\ref{Fig:X1.0} and~\ref{Fig:X1.5},
we show the dependence of the ratio $\chi_4/\chi_2$ on the temperature 
for different system sizes and different anisotropy parameters. The calculations 
are done  on lines of constant ratio of $\mu/T$. We consider the values
$\mu/T = 0, 0.5, 1, 1.5$. 
We observe that the cumulant ratio does not vary much at high temperatures $T>1.6 T_{\rm pc}$
and is almost 
independent of the system size. 
However there is a significant variation in the vicinity of the phase transition and 
at lower temperatures. 
The figures also show that the location of the maximum of $\chi_4/\chi_2$ 
shifts to lower temperatures with the decreasing system size.   
The behavior of the maximal value of  $\chi_4/\chi_2$
on the system size is non-monotonous: with decreasing $L$,
the maximum first decreases until $L$ reaches about 3 fm 
and than increases. 
The dependence on the system size becomes more complicated 
at higher chemical potential, because  the cumulants 
become sensitive to the ACP II.

\begin{figure}
\includegraphics[width=0.325\linewidth]{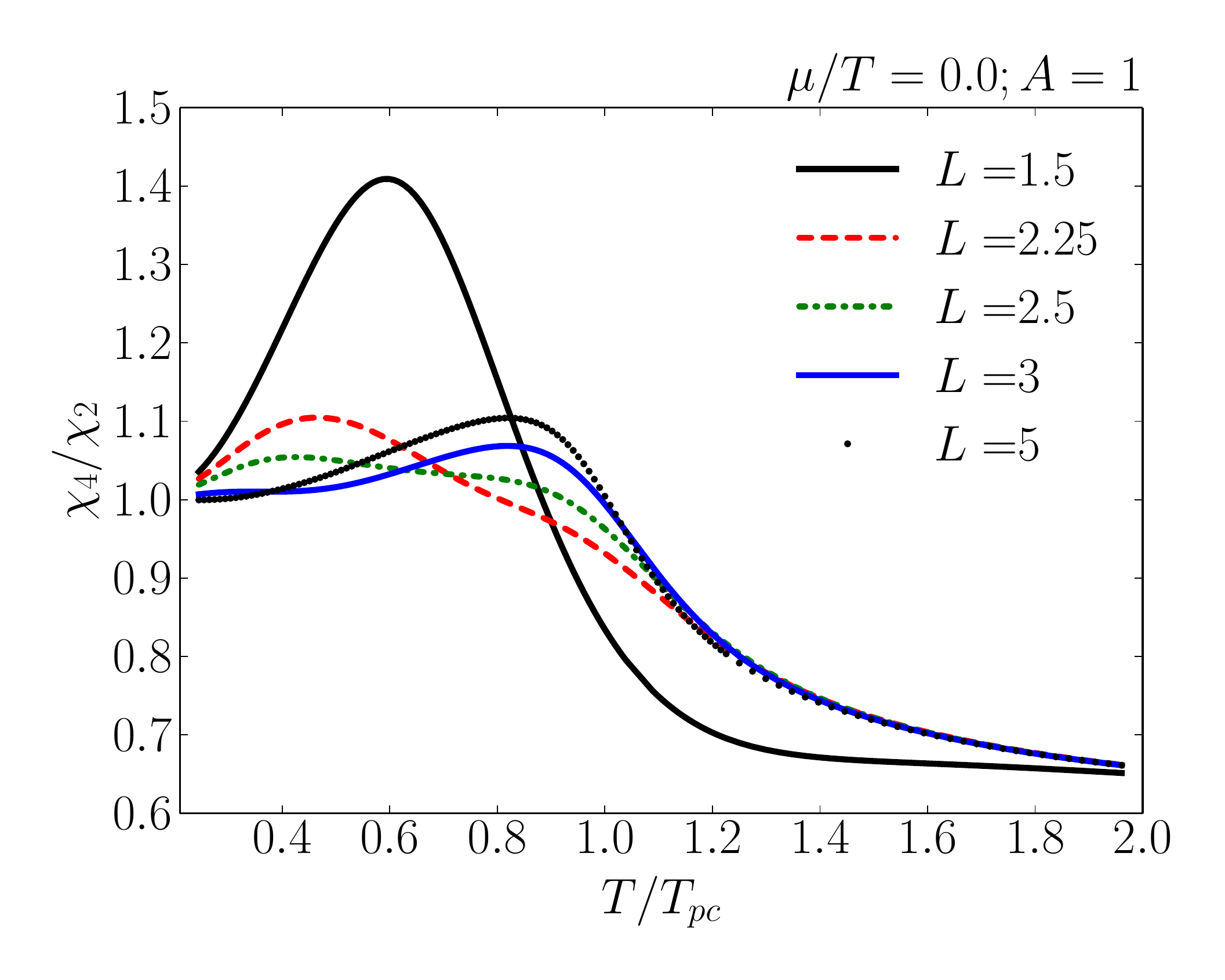}
\includegraphics[width=0.325\linewidth]{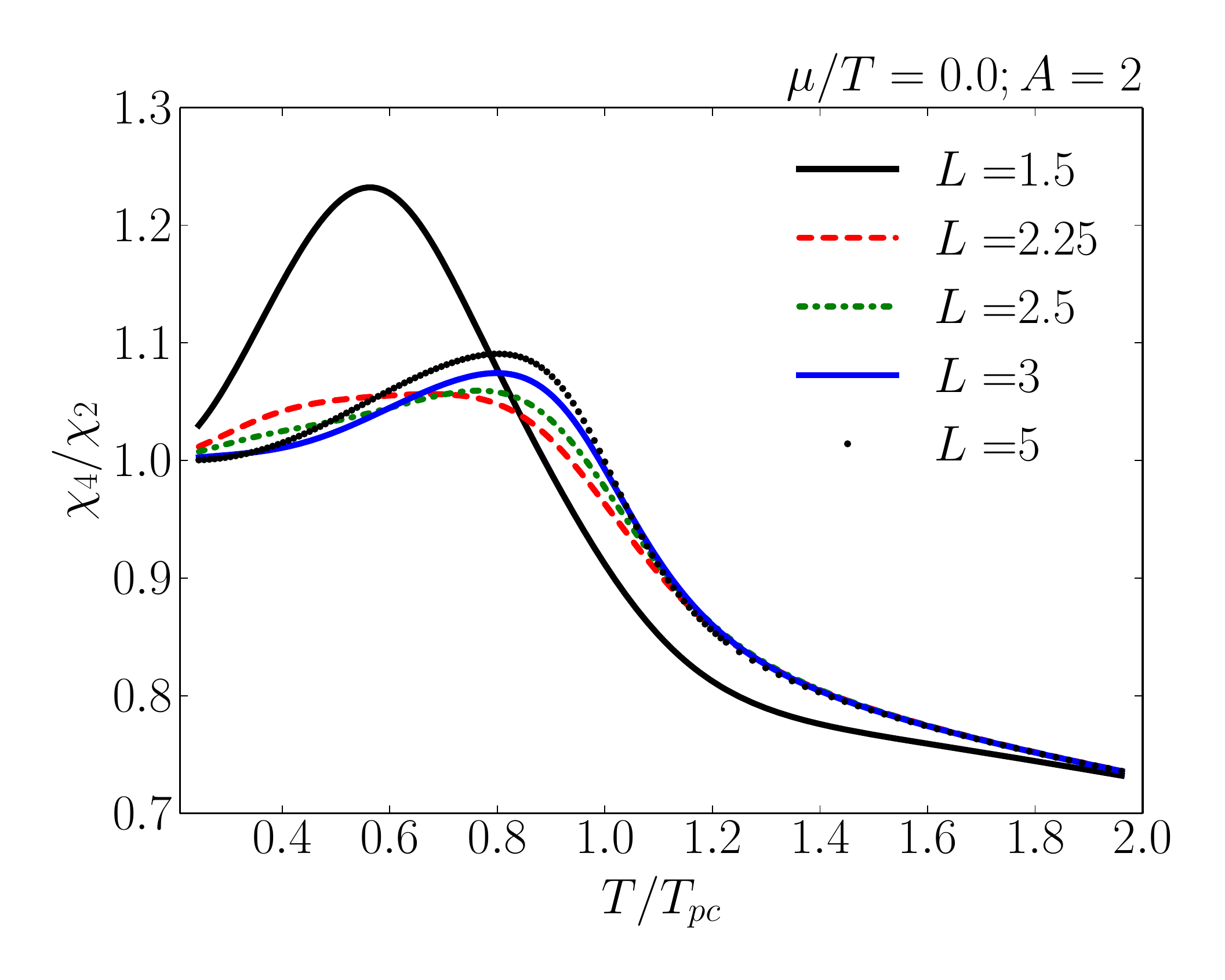}
\includegraphics[width=0.325\linewidth]{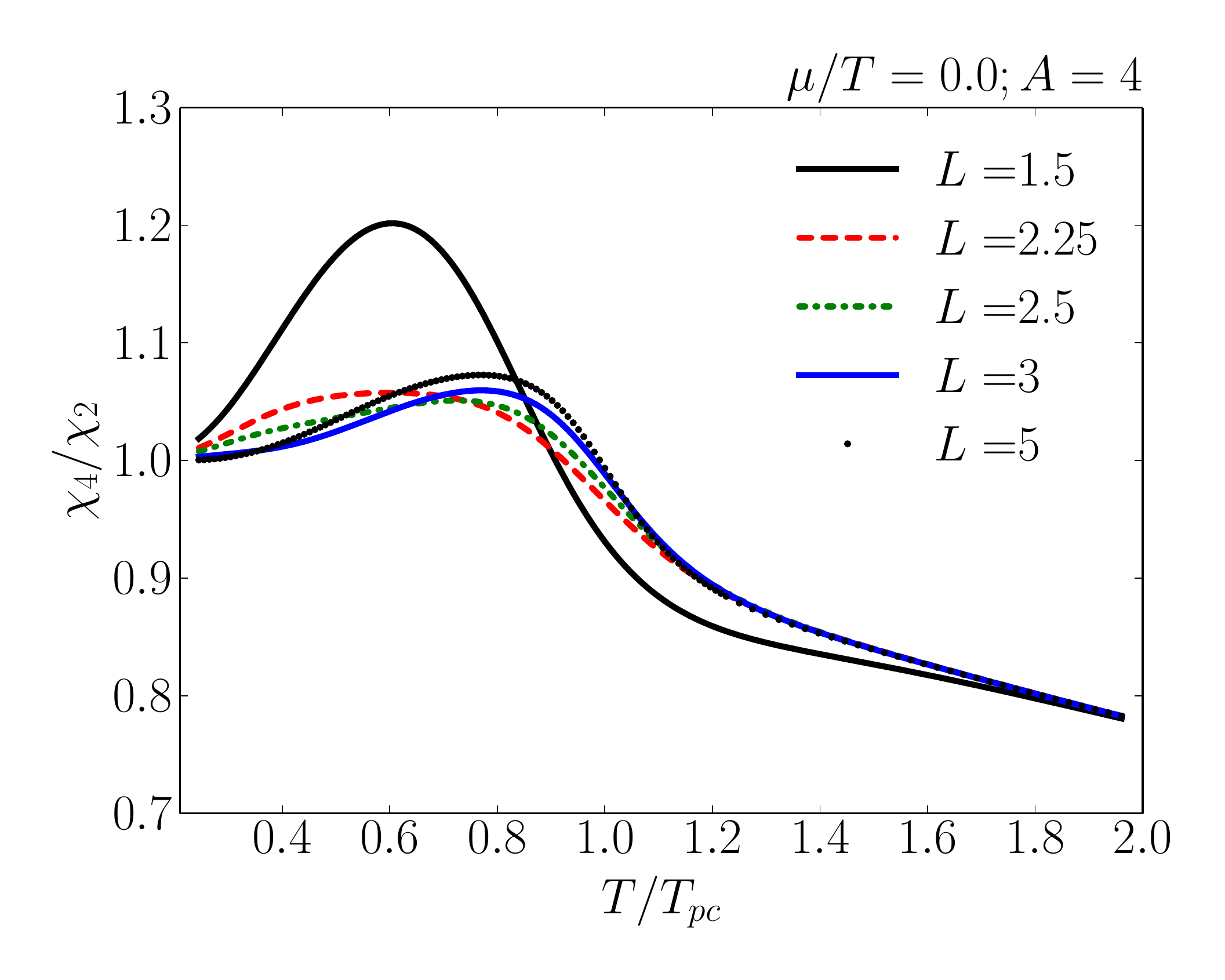}
\caption{
	The ratio of the fourth to the second order susceptibilities 
	as a function of temperature for different systems sizes 
	and the anisotropy parameter $A$; the results are computed at 
	zero chemical potential. 
}
\label{Fig:X0}
\end{figure}

\begin{figure}
	\includegraphics[width=0.325\linewidth]{{{test_0.5_L}}}
	\includegraphics[width=0.325\linewidth]{{{test_0.5_as2_L}}}
	\includegraphics[width=0.325\linewidth]{{{test_0.5_as4_L}}}
\caption{
	The ratio of the fourth to the second order susceptibilities 
	as a function of temperature for different systems sizes 
	and the anisotropy parameter $A$; the results are computed at 
	$\mu/T=0.5$. 
}
\label{Fig:X0.5}
\end{figure}

\begin{figure}
	\includegraphics[width=0.325\linewidth]{{{test_1.0_L}}}
	\includegraphics[width=0.325\linewidth]{{{test_1.0_as2_L}}}
	\includegraphics[width=0.325\linewidth]{{{test_1.0_as4_L}}}
\caption{
	The ratio of the fourth to the second order susceptibilities 
	as a function of temperature for different systems sizes 
	and the anisotropy parameter $A$; the results are computed at 
	$\mu/T=1$. 
}
\label{Fig:X1.0}
\end{figure}

\section{Conclusions}

In this article, we considered the quark-meson model in a finite volume. We carried out our calculations using the functional renormalization group approach. We demonstrated that the previously employed Litim regulator is not suitable for finite volume studies in small values, and we proposed to use an exponential regulator instead.

We computed the chiral susceptibility on the phase diagram in finite volume and we showed that for some volumes there are two distinct apparent critical end points. 
One of them, which we called ACP I, is smoothly connected to the critical point of the infinite volume calculation when its location is considered in the function of volume. The location of this point moves to lower temperatures and higher chemical potentials with decreasing system size. The other apparent critical endpoint, ACP II, approaches the zero temperature axis and is not detectable for system size 
larger than 4 fm. For small system sizes however its location shifts towards 
higher temperatures and lower chemical potentials with decreasing system size.

Our main goal was to calculate the ratio of the fourth to the second order baryon number cumulant. 
These calculations showed that there is a rather strong volume dependence of the ratio
for system size less than 5 fm. This dependence becomes more significant with increasing 
value of chemical potential, because it probes regions of the phase diagram which are
close to the apparent critical point ACP II.

Our results indicate that an estimate of the effect of the volume 
fluctuations~\cite{Skokov:2012ds,Braun-Munzinger:2016yjz}   for system sizes less than 5 fm  
might be very challenging and should account not only 
for the explicit, but also for the implicit volume dependence of the cumulants.

\appendix 
\section{Mode summation}
\label{Sec:ModeSum}
In this appendix, we consider an efficient  numerical way to perform summation of the discrete modes  for 
the different boundary conditions.

\subsection{Isotropic periodic boundary conditions} 
\label{Sect:IPBC} 
The calculation in a finite volume with the periodic boundary conditions 
involves a three dimensional summation of 
functions that depend only on the magnitude 
$\vec{n}^2 = n_x^2 + n_y^2 + n_z^2$.  We use this symmetry to rewrite the sum 
as
\begin{equation}
\sum_{\vec{n}} f(\vec{n}^2)  = 
\sum_{m=0}^{\infty} \left( \sum_{\vec{n}}  \delta_{m,\vec{n}^2} \right) 
f (m) = \sum_{m=0}^{\infty} G(m) f(m),
\label{Eq:FSUM}
\end{equation}
where the function, the multiplicity of states,  $G(m)\equiv \sum_{\vec{n}}  \delta_{m,\vec{n}^2} $ 
automatically  takes the symmetries of the  magnitude  
$\vec{n}^2 = n_x^2 + n_y^2 + n_z^2$ into account.  The function $G(m)$ is 
to be computed once and tabulated for repeated use. This method gives a significant 
reduction in computational time.

\subsection{Anisotropic periodic boundary conditions} 
Here, following the logic of Sec.~\ref{Sect:IPBC}, we extend   
the method for anisotropic volumes. Let us consider particular anisotropy 
$L_z = A L_x = A L_y = A L$, where $A$ is an integer number.
The momentum is, just as in the isotropic case, discrete and its magnitude is given by 
\begin{equation}
	\vec{p}^{\;2} = \left( \frac{2 \pi}{L A} \right)^2 \left( A^2 n_x^2 + A^2 n_y^2 +  n_z^2 \right).  
\end{equation}
Thus, as before, we can introduce the multiplicity of states 
\begin{equation}
	G^A(m) = \sum_{\vec{n}} \delta_{m, A^2 n_x^2+ A^2 n_y^2+ n_z^2} 
\end{equation}
to perform the summation
\begin{equation}
	\sum_{\vec{n}} f(A^2 n_x^2+A^2 n_y^2+n_z^2) =  \sum_{m=0}^{\infty} G^A(m) f(m)\, .
\end{equation}

\subsection{Anti-periodic boundary conditions}
Although not used in this paper, for completeness  we also consider the 
anti-periodic boundary conditions
\begin{align}
	\vec{p}^{\; 2}  &=  \left( \frac{ \pi}{L} \right)^2 \left(  (2n_x+1)^2 + (2n_y+1)^2 +  (2n_z+1)^2 \right) 
	\notag \\ &= 
 	\left( \frac{ 2 \pi}{L} \right)^2 
	\left( n_x^2+n_x + n_y^2 + n_y  + n_z^2 +n_z + \frac34 \right) . 
\end{align}
Thus it is convenient to introduce 
\begin{equation}
	G^{\rm AP}(m) = \sum_{\vec{n}} \delta_{m,  n_x^2 + n_x + n_y^2 + n_y + n_z^2 + n_z} 
\end{equation}
so that summation can be represented as 
\begin{equation}
	\sum_{\vec{n}} f( (2n_x+1)^2 + (2n_y+1)^2 +  (2n_z+1)^2  ) =  
	\sum_{m=0}^{\infty} G^{\rm AP}(m) f(4m+3)\, .
\end{equation}

\section{Numerical details and tests}
\label{Sec:Num}
In order to explore the region of the phase diagram at a 
high chemical potential, where $\Omega_k(\sigma)$  may potentially
develop two minima, we avoided the polynomial/Taylor expansion anzatz for
the thermodynamic potential. An alternative approach would be to use the 
so-called grid method; which is based on the evenly spaced discretization  
of the classical field, $\sigma$, see e.g.~Ref.~\cite{Schaefer:2004en}. It is however very well 
known that uniform discretization results in the worse possible 
approximation of a function. Instead we use the pseudo-spectral
Chebyshev collocation method. We found that this method is numerically more
reliable and substantially faster than the grid method. The details of the 
numerical method can be found in Ref.~\cite{Borchardt:2016pif,AlmSko2016};
here we only present 
the most important ingredients. 
The function $\Omega_k(\sigma)$ is approximated by the Chebyshev expansion up to the $N$-th order
\begin{equation}
	\Omega_k(\sigma) = \sum_{i=0}^{N-1} a_i(k) T_i(\sigma)\;.  
\end{equation}
The flow equation is then solved in the range
$-\sigma_{\rm max} < \sigma < \sigma_{\rm max} $
at the collocation nodes defined by the zeros
of $T_N(\sigma)$. The boundary conditions have to be provided additionally. 
To enhance stability, we keep the meson masses at the outermost collocation points constant during the flow given by their UV value. The maximal value of the field, $\sigma_{\rm max} = 400$ MeV, 
and the order of Chebyshev approximation, $N=120$,   were chosen by testing
the convergence of the results (obviously, physics should be 
independent of our choice of either $\sigma_{\rm max}$ or $N$). 

We note that it is absolutely crucial to use $\sigma_{\rm max} \geq 400$ MeV,
at or below  $L=1$ fm.  
While the final value of $\sigma$ (minimum of the potential) is only a few dozen percents above $f_\pi$; 
it may reach substantially larger values, $\sim 300$ MeV at intermediate  $k$. 

The free parameters and the initial conditions are defined to describe the 
following vacuum properties in the infinite system  
\begin{itemize}
	\item the pion decay constant, $f_\pi= 93 $ MeV, 
	\item the constituent quark mass, $m_q = 300$ MeV,  
	\item the pion mass, $m_\pi = 140$ MeV, 
	\item the sigma mass, $m_\sigma = 585$ MeV.
\end{itemize}
These result in 
$y = 3.2$, and $\Omega(k=\Lambda, \sigma) = m_I^2 \frac{\sigma^2}{2} +  \lambda_I \frac{\sigma^4}{4}$ with 
$m_I = 753$ MeV and $\lambda_I = 27.8$, together with the UV cutoff $\Lambda = 950$ MeV. 
\section{Absence of spontaneous symmetry breaking in finite volume}
\label{Sec:ASB}
In this section, we discuss the absence of spontaneous symmetry breaking in 
finite volume with periodic boundary conditions. To this end, we consider the quark-meson model without explicit chiral symmetry breaking, so the external field is set to $h=0$.  

The periodic boundary conditions naturally include a zero mode $p_x=p_y=p_z=0$. 
Therefore, as in the case of the infinite volume, we expect the potential
to evolve to a convex one, in contrast to anti-periodic boundary conditions, which do {\it not} include the zero mode. Using the numerical calculations of a finite volume system, we want to demonstrate that in this case the spontaneous symmetry 
breaking is impossible. We will start, however, 
with an analytic argument. Let us consider a schematic form of the flow equation:
\begin{equation}
	\partial_k \Omega_k \propto \frac{1}{L^3} \frac{k T}{k^2  +  m_k^2} \,, 
\end{equation}
where we considered sigma meson contribution and  included the $\vec{n}=0$ 
zero mode. Higher modes do not play an important role at small $k$, where the most important 
part of the FRG evolution takes place. Other degrees of freedom can be included as well, but 
they do not change the main conclusion. Additionally, we only consider the ``high-temperature'' limit of the 
Bose-Einstein distribution function, i.e. we approximate 
\begin{equation}
	1 + 2 n_B(\omega) \approx \frac{2 T}{\omega}\, .   
\end{equation}
This is a good approximation for the zero mode at $k \ll T $ and $m_k  \ll T $. These restrictions are suitable to study the spontaneous symmetry breaking.
\begin{figure}
\includegraphics[width=0.7\linewidth]{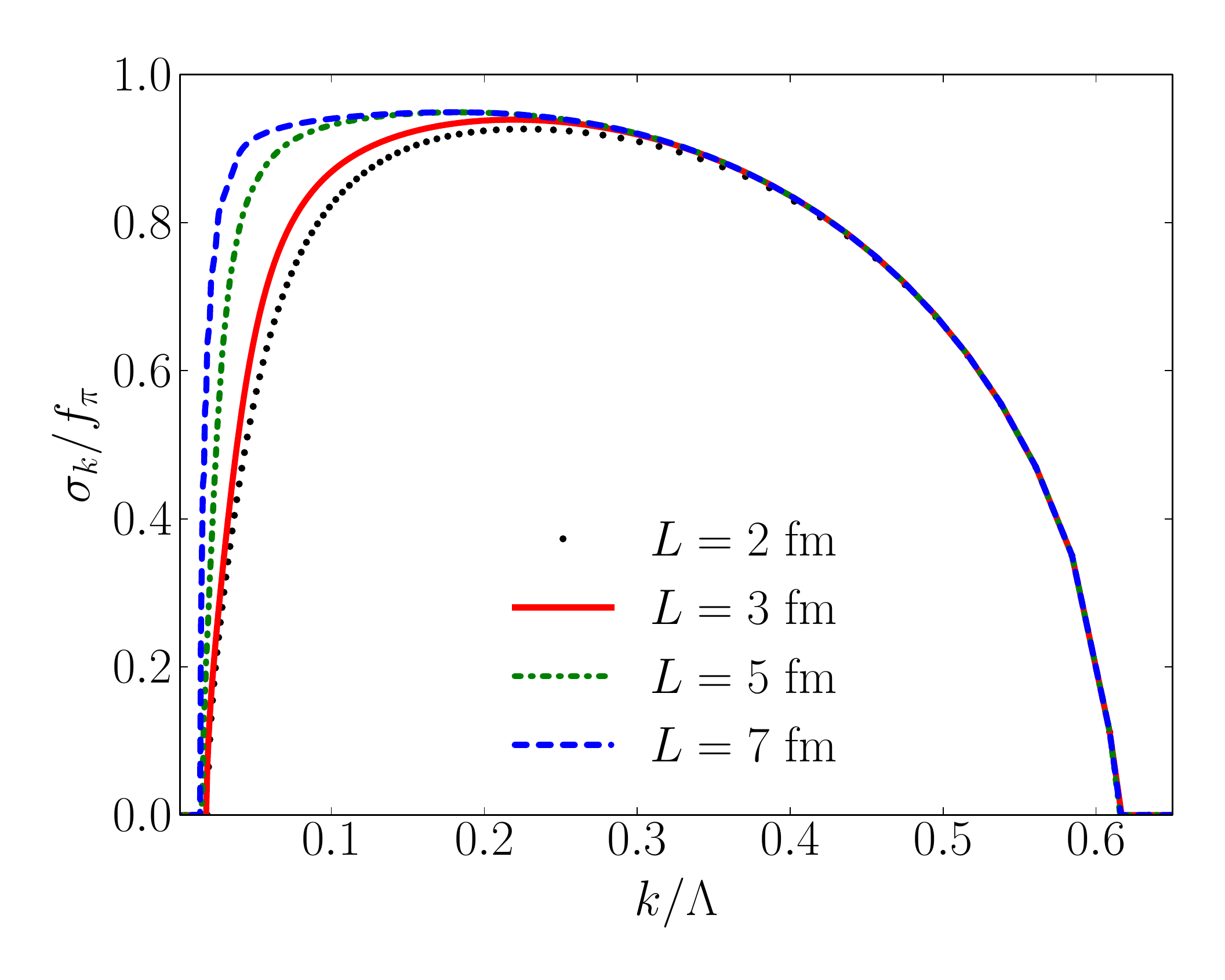}
\caption{The dependence of the minimum of the potential 
	on the FRG flow parameter $k$ for different system sizes in  the chiral limit $h=0$. The calculations are performed at $T = $ 10 MeV. For higher $L$, the non-trivial minimum of the potential is preserved by the FRG evolution to lower values of $k$, as expected.   }
\label{Fig:Sigma_k_SSB}
\end{figure}

Next we assume that $ \lim_{k\to 0} \partial_k \Omega_k = 0$. This condition manifests the convergence
of the FRG flow equation. It also implies that 
\begin{equation}
	\lim_{k\to0}  \frac{k T}{k^2  +  m_k^2} = 0\, ,
\end{equation}
which has a few important consequences. First of all, it restricts $m_k^2$ from being non-negative 
to positive values only. Next, it also demands that at small 
$k$, $m_{k}^2$ cannot be proportional to a larger or equal positive power of $k$ to unity. That being said, the masses, and hence the curvature will either converge to a positive constant, in which case there is no spontaneous symmetry breaking, or go to zero with a power of $k$ smaller than one. In the latter case however, since $m_{k}^2$ approaches zero slower, than $k^2$, the bosonic modes will decouple. The dynamics is then purely fermionic, hence Landau-treatment is possible. The resulting potential will be analytic, and this rules out the possibility of $m_{k}^2$ approaching zero with a power of $k$ between 0 and 1. This rules out the second possibility, so $m_{k}^2$ has to approach a positive constant, no spontaneous symmetry breaking possible. Following similar logic, this argument can be also easily extended to the $T=0$ limit; we leave this as an exercise for an interested reader. 

Note that the above argument does not restrict $m_k^2$ being negative  at some non-zero $k$. Indeed our 
numerical simulation do show that the transitional potential does develop a minimum at some non-zero 
$\sigma$; it, however, evolves at small $k$ to $\sigma=0$. Intuitively the transition between this two 
regimes starts at the values of $k$ inversely proportional to the system size, $L$. We confirmed this
with direct numerical calculations shown in Fig.~\ref{Fig:Sigma_k_SSB}.    

As seen in Fig.~\ref{Fig:Sigma_k_SSB}, 
the initial evolution of the minimum is independent of the system size. The curves start to deviate 
from each other when the discreetness of the momentum start to play an important role for mesonic 
fluctuations.

\section{Perturbative contribution}
\label{Sec:Pert}
As we alluded to in Section \ref{Sec:Cumulants}, the contribution of quarks is not negligible above the cutoff $\Lambda$
and must be properly accounted for. Here we follow  Ref.~\cite{Skokov:2010wb}
and supplement the FRG flow above the cutoff with the following contribution 
\begin{equation}
\label{eq:gen_floweq_fin_3}
	\partial_k \Omega(k>\Lambda) = \frac{-2\nu_q }{4L^3} \sum_{n_x,n_y,n_z} \frac{1-n_F(E_q)-\bar{n}_F(E_q)}{E_q}  \partial_k R_k(q),
\end{equation}
where $E_q$ is computed using the perturbative quark mass $m_q=0$. 
The vacuum contribution here can be neglected because it does not 
alter the dynamics. 
This equation can be integrated out to yield:
\begin{align}
	\Delta \Omega &= \int_{\infty}^{\Lambda} \partial_k \Omega_q = \int^{\infty}_{\Lambda}
	\frac{\nu_q}{2 L^3}  \sum_{n_x,n_y,n_z} 
	\left( \frac{n_F(E_q)+\bar{n}_F(E_q)}{E_q}\right) \partial_k R_k(q)dk \\
	&= -\frac{T\nu_q}{L^3}  
	 \sum_{n_x,n_y,n_z} 
	\left( \log\left(1+e^{\frac{\mu-E_q^{\Lambda}}{T}}\right) + \log\left(1+e^{-\frac{\mu+E_q^{\Lambda}}{T}}\right)\right), 
\end{align}
where 
$ E_q^\Lambda  = \sqrt{m_q^2 + q^2 + R_\Lambda(q)}$.

\section{Large/small $L$ limits in the mean-field approximation}
\label{Sec:Limits} 
To get the mean-field (MF) approximation we start from the RG flow equation in finite volume and we drop the boson contribution. This yields
\begin{equation}
\label{eq:gen_floweq_fin_2}
	\partial_k \Omega = -\frac{\nu_q}{2L^3} \sum_{n_x,n_y,n_z}
	\left(\frac{1-n_F(E_q)-\bar{n}_F(E_q)}{E_q}\right) \partial_k R_k(q)
\end{equation}
for an arbitrary regulator function $R_k(q)$. 
 The quark energies are given by
\begin{equation}
	E_q= \sqrt{g^2\sigma^2 + q^2+R_k(q)},\quad q =\frac{2\pi}{L} \sqrt{n_x^2+n_y^2+n_z^2}.
\end{equation}
For simplicity let us consider $T=\mu=0$ to understand the asymptotic behavior of the theory in the function of system size. The grand canonical potential is given by
\begin{align}
	\Omega &= U_{\Lambda}(\sigma) + \int_{\Lambda}^0 dk \partial_k \Omega \\
	&= U_{\Lambda}(\sigma) + \frac{\nu_q}{2L^3}\int_0^{\Lambda}  \sum_{n_x,n_y,n_z}
	\frac{\partial_k R_k(q)}{E_q}dk = U_{\Lambda}(\sigma) 
	+ \sum_{n_x,n_y,n_z} \frac{\nu_q}{L^3}\int_0^{\Lambda} \frac{dE_q}{dk}dk \\
	&= U_{\Lambda}(\sigma) + \frac{\nu_q}{L^3}\sum_{n_x,n_y,n_z} \left(\sqrt{g^2\sigma^2+q^2+R_{\Lambda}(q)}-\sqrt{g^2\sigma^2+q^2}\right).
\end{align}
The first term is highly suppressed for small momenta, however is essential for the UV regularization of the theory.

\subsection{Small L behavior}
If $L\rightarrow 0$  all momenta will be large, except the zero mode. The contribution to the sum will be dominated by the zero mode contribution, that depends on $L$ as $L^{-3}$:
\begin{equation}
	\Omega_{0} = \nu_q\frac{\sqrt{g^2\sigma^2+R_{\Lambda}(0)}- |g\sigma|}{L^3}.
\end{equation}
 Apart from the zero mode all other mode will have diverging momentum as $L\rightarrow 0$. The contribution of a mode can be obtained using the approximation
\begin{equation}
	\sqrt{m^2+q^2} = q\left(1+\frac{m^2}{2q^2}-\frac{m^4}{8q^4}\right)+\ldots
\end{equation}
and yields
\begin{equation}
 	\Omega_{n} = \frac{\nu_q R_{\Lambda}(q_n)}{4\pi L^2} - \nu_q\frac{2g^2\sigma^2 R_{\Lambda}(q_n)+R_{\Lambda}(q_n)^2}{64\pi^3}. 
\end{equation}
In the case of proper UV regularization these contributions vanish for $L\rightarrow 0$. One can see that the zero mode contribution alone cannot fulfill the gap equation, and tries to push the condensate to $\sigma\rightarrow \infty$. If we neglect the other modes (due to exact cancellation or exponential suppression above the UV cutoff), then at finite $L$ an interplay of the zero mode and the UV potential $U_{\Lambda}(\sigma)$ will yield the condensate and hence it is expected that it increases with a power law. In particular with the exponential regulator, at $L\rightarrow 0$, assuming a $\lambda/4 \sigma^4$ leading term in $U_{\Lambda}(\sigma)$, the gap equation will asymptotically be
\begin{equation}
	\lambda \sigma^3 = \frac{\nu_q}{2L^3} \frac{\Lambda^2}{g\sigma^2},
\end{equation}
yielding $\sigma \sim L^{-3/5}$.

\subsection{Large L behavior}
The large $L$ behavior will be opposite to the small $L$ behavior, here the interplay of many modes will yield the final result. We use the Poisson-summation method to obtain the result. We start from the identity
\begin{equation}
	\sum_{k=-\infty}^{\infty} e^{-2\pi i k x} = \ldots+\delta(x-2)+\delta(x-1)+\delta(x)+\delta(x+1)+\delta(x+2)+\ldots
\end{equation}
which can be used to yield
\begin{align}
\sum_{n=-\infty}^{\infty} f(n) &= \int_{-\infty}^{\infty} dx f(x) \left(\ldots+\delta(x-2)+\delta(x-1)+\delta(x)+\delta(x+1)+\delta(x+2)+\ldots\right) \\
&= \int_{-\infty}^{\infty} dx f(x)\sum_{k=-\infty}^{\infty} e^{-2\pi i k x}
= \sum_{k=-\infty}^{\infty}\int_{-\infty}^{\infty} dx f(x) e^{-2\pi i k x}.
\end{align}
In the finite size setup we have summation over momentum modes in 3 dimensions. In all three we apply this identity and change variable from the discrete mode number back to the physical momenta. This formally in the $x$ direction is
\begin{equation}
	\frac{1}{L} \sum_{n_x=-\infty}^{\infty} f\left(\frac{2\pi n_x}{L}\right)
	= \sum_{j=-\infty}^{\infty}\int_{-\infty}^{\infty} \frac{dq}{2\pi} f(q) e^{-i j q L}.
\end{equation}
Applying this in all direction yields for the grand canonical potential
\begin{equation}
	\Omega = U_{\Lambda}(\sigma) + \nu_q\sum_{j_x,j_y,j_z} \int \frac{d^3q}{(2\pi)^3} \left(\sqrt{g^2\sigma^2+q^2+R_{\Lambda}(q)}-\sqrt{g^2\sigma^2+q^2}\right) e^{-i L \vec{j}\vec{q}}.
\end{equation}
In this sum we can consider only the smallest winding numbers, $j_i = \pm 1$;  
and perform integration using the saddle point approximation. 
This is a straightforward derivation and as such requires only brief description. 
Let us consider only the winding number $j_z=1$. The integration with respect to 
the transverse coordinate $q^2_\perp = q_x^2 + q_y^2$ can be performed 
analytically assuming that $q_\perp \ll \Lambda$. This assumption is justified 
because only the momenta $q<\Lambda$ contribute to the integral. 
The upper bound of the integration is thus also limited by $\Lambda$. 
After the integration is performed, one can use saddle point approximation
to integrate with respect to $q_z$. We obtain that the 
correction to the infinite volume limit, $\vec{j}=0$, is proportional to 
$\frac{g \sigma \Lambda}{L^2} \exp (- g\sigma L)$. 
Hence it is expected that at the $L\rightarrow \infty$ limit the order parameter will approach the infinite volume value exponentially with the system size $L$.

\section*{Acknowledgments}
We thank B.~Friman, S.~Rechenberger, K.~Redlich,  S.~Mukherjee for useful discussions. 
We are grateful to A. Bzdak for comments and suggestions. R.D.P. thanks the U.S. 
Department of Energy for support under contract DE-SC0012704. 

\bibliography{fs}

\newpage

U.S. Department of Energy Office of Nuclear Physics or High Energy Physics

{\it Notice:} 
This manuscript has been co-authored by employees of Brookhaven 
Science Associates, LLC under Contract No. DE-SC0012704 with 
the U.S. Department of Energy. The publisher by accepting the manuscript for 
publication acknowledges that the United States Government retains a 
non-exclusive, paid-up, irrevocable, world-wide license to publish or 
reproduce the published form of this manuscript, or allow others to do so, 
for United States Government purposes.
This preprint is intended for publication in a journal or proceedings.  
Since changes may be made before publication, it may not be cited or 
reproduced without the author’s permission.

{\it DISCLAIMER}:
This report was prepared as an account of work sponsored by an agency of the 
United States Government.  Neither the United States Government nor any 
agency thereof, nor any of their employees, nor any of their contractors, 
subcontractors, or their employees, makes any warranty, express or implied, 
or assumes any legal liability or responsibility for the accuracy, 
completeness, or any third party’s use or the results of such use of any 
information, apparatus, product, or process disclosed, or represents that 
its use would not infringe privately owned rights. Reference herein to any 
specific commercial product, process, or service by trade name, trademark, 
manufacturer, or otherwise, does not necessarily constitute or imply its 
endorsement, recommendation, or favoring by the United States Government or 
any agency thereof or its contractors or subcontractors.  The views and 
opinions of authors expressed herein do not necessarily state or reflect 
those of the United States Government or any agency thereof. 

\end{document}